\documentclass[aps,prd,12pt,showpacs,groupedaddress,preprintnumbers,floatfix,nofootinbib]{revtex4}

\usepackage{amsmath}
\usepackage{graphicx}
\usepackage{color}
\usepackage{dcolumn}% Align table columns on decimal point
\usepackage{bm}% bold math

\begin{document}

\preprint{DCP-09-02}

%opening
\title{Anomalies, Beta Functions and Supersymmetric Unification with Multi-Dimensional 
Higgs Representations}

\author{
  Alfredo Aranda,$^{1,2}$\footnote{Electronic address:fefo@ucol.mx}
  J. L. D\'iaz-Cruz,$^{2,3}$\footnote{Electronic address:ldiaz@sirio.ifuap.buap.mx}
  and Alma D. Rojas$^{3}$}

\affiliation{$^1$Facultad de Ciencias and CUICBAS \\  Universidad de Colima, Colima, Col. M\'exico\\
  $^2$Dual C-P Institute of High Energy Physics \\
  $^3$C.A. de Part\'iculas, Campos y Relatividad \\
  FCFM-BUAP, Puebla, Pue. M\'exico}

\date{\today}

\begin{abstract}
  In the framework of supersymmetric Grand Unified Theories, the minimal Higgs sector is often 
extended by introducing multi-dimensional Higgs representations in order to obtain realistic models. 
However these constructions should remain anomaly-free, which constraints significantly their
structure. We review the necessary conditions for the cancellation of anomalies in general
and discuss in detail the different possibilities for SUSY SU(5) models.
Alternative anomaly free combinations of Higgs representations, beyond the
usual vector-like choice, are identified, and it is shown that their corresponding 
$\beta$ functions are not equivalent. Although the unification of gauge couplings is not 
affected, the introduction of multi-dimensional representations leads to different scenarios 
for the perturbative validity of the theory up to the Planck scale.
\end{abstract}


\maketitle

\newpage

\section{Introduction}
Supersymmetric Grand Unified Theories
(SGUT)~\cite{GeorgiGlashow74,Raby,Mohapatra,Raby2} have achieved some
degree of success: unification of gauge couplings, charge
quantization, prediction of the weak mixing angle, the mass-scale
of neutrinos. Detection of weak scale superpartners or proton
decay, as well as some patterns of FCNC/LFV and CP violation
phenomena would indicate that some form of SGUT lays beyond the
SM~\cite{JLDC}. Although this degree of success is already present
in the minimal models (SO(10) or some variant of
SU(5))~\cite{Murayama,HMY93,Barr}, there are open problems that suggest
the need to incorporate more elaborate constructions~\cite{HMTY},
specifically the use of higher-dimensional representations in the
Higgs sector (i.e. SU(5) representations with dimension $> 24$). 
For instance, a \textbf{45} representation is often
included to obtain correct mass relations for the first and second
families of d-type quarks and leptons~\cite{GeorgiJarlskog}, while
a \textbf{75} representation has been employed to address the
doublet-triplet problem~\cite{MuOkYa}.

When one adds these higher-dimensional Higgs representation within
the context of ${\cal N}=1$ SUSY GUTs, one must verify the
cancellation of anomalies associated to their fermionic partners,
i.e. the Higgsinos. The most straightforward solution to
anomaly-cancellation is obtained by including vector-like
representations i.e. including both $\psi$ and $\bar{\psi}$ chiral
supermultiplets; up to our knowledge this seems to be the option
chosen by most models builders. It is one of the purposes of this
paper to find alternatives to this option, namely to create an
anomaly free Higgs sector, including some representation $\psi$
and a set of other representations of lower-dimension
$\left\lbrace \phi_{1},\phi_{2},...\right\rbrace $. It turns out
that different anomaly-free combinations of representations are not
equivalent in terms of their $\beta$~functions.

It is also known that the unification condition imposes some
restrictions on the GUT-scale masses of the gauge bosons,
gauginos, Higgses, and  Higgsinos~\cite{Terning2}. However the
addition of complete GUT multiplets does not change the unified
gauge couplings, and neither modifies the unification scale.
 On the other hand, the
evolution of the gauge couplings above the GUT scale, up to the
Planck scale, depends on the matter and Higgs content, thus the 
perturbative validity of the model is affected by
the inclusion of additional multiplets. This is important in order
to determine  whether gravitational effects  should be invoked for
the viability of the model~\cite{Calmet}. In this paper we also
study the effect of the higher-dimensional Higgs multiplets on
the evolution of the gauge coupling up to the Planck scale, 
focusing on  models that invoke different sets of
representations in order to satisfy the anomaly-free
conditions.

Our paper is organized as follows. In section II we review the different
mechanisms proposed in the literature to get anomaly free gauge
theories for general gauge groups. Focusing on SU(N)-type models,
we look for new alternatives to anomaly cancellation. The implications of our results 
for specific SUSY GUT SU(5) models
are presented in section III. The issue of gauge coupling unification,
and the effect of higher-dimensional representations is discussed
in section IV, where we include the 2-loop effect on the gauge unification
that is brought by the Yukawa couplings associated with those representations
at the 1-loop level. 
Finally our conclusions are presented in section V.

\section{Anomalies in gauge theories}
Whether a symmetry that holds at the classical level is respected
or not at the quantum level is signaled by the presence of
anomalies. The importance of anomalies was recognized almost
immediately after the proof that Yang-Mills theories with SSB are
renormalizable was presented in~\cite{Hooft Veltman}. Anomalies can be associated
with both global or local symmetries, the latter being most
dangerous for the consistency of the theory. The so-called
perturbative anomalies arise in abelian gauge symmetries while
non-abelian symmetries can have anomalies of a non-perturbative
origin that turn out to be of topological
nature~\cite{AdlerBellJackiw}. The need to require anomaly
cancellation in any gauge theory stems from the fact that their
presence destroys the quantum consistency of the
theory~\cite{Bardeen}. It turns out that all one needs in order
to identify the anomaly is to calculate the triangle diagrams
of the form AVV, with A=Axial, and V=Vector currents.

For a given fermionic representation $D$ of a gauge group $G$, the
anomaly can then be written as~\cite{Terning2}:
\begin{equation}
  A(D)d^{abc}\equiv Tr\left[ \left\lbrace  T_a^{D_i},T_b^{D_i}\right\rbrace T_c^{D_i} \right] ,
\end{equation}
where $T_a^{D_i}$ denotes the generators of the gauge group $G$ in
the representation $D_i$, and $d^{abc}$ denotes the anomaly
associated with the fundamental representation. The anomaly
coefficients $a_D\equiv A(D)$ for the most common representations
are shown in Table~\ref{table:SUNreps1} for SU(N) groups; a result
that is known in the literature~\cite{Terning2}.  In order to
obtain these results one makes use of the following relations:

\begin{enumerate}
    \item For a representation $R$ that is a direct sum of two
    representations, $R=R^{1}\oplus R^{2}$, the anomaly is given
    by
    \begin{equation}\label{directsum}
    A_{R}=A(R_{1}\oplus R_{2})=A(R_{1})+A(R_{2}).
\end{equation}
    \item For a representation $R$ that is the tensor product of two
    representations, the anomaly is given by:

    \begin{equation}\label{tensorproduct}
    A_{R}=A(R)=A(R_{1} \otimes R_{2})=D(R_{1})A(R_{2})+
    D(R_{2})A(R_1),
     \end{equation}
     with $D(R_{i})$ denoting the dimensions of representations
    $R_{i}$.
\end{enumerate}
Then, starting from the fundamental representations $F$, we
have taken the tensor products and evaluated the
unknown coefficients that appear in the products in terms of
$A(F)$. The dimension of the representations has been verified
using the chain notation $(\alpha,\beta,\gamma,\ldots)$.
Results for some SU(N) representations can be read off from tables
in~\cite{Slansky}. We have extended these results to include 
additional higher-dimensional representations, with the corresponding
expressions shown in Table~\ref{table:SUNreps2}.

\begin{table}[ht]
\[\begin{array}{|c|c|c|c|}\hline
 Irrep & dim(r)& 2T(r) & A(r)\\  \hline
 \includegraphics[scale=.15]{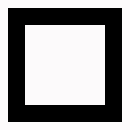}&N&1&1\\
  Ad &N^{2}-1&2N&0\\
    \includegraphics[scale=.15]{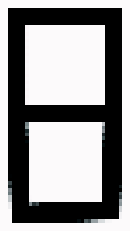}&\frac{N(N-1)}{2}&N-2&N-4\\
  \includegraphics[scale=.15]{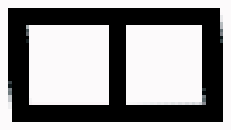}&\frac{N(N+1)}{2}&N+2&N+4\\
  \includegraphics[scale=.15]{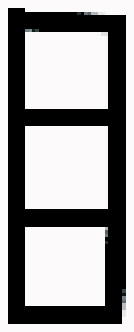}&\frac{N(N-1)(N-2)}{6}&\frac{(N-3)(N-2)}{2}&\frac{(N-3)(N-6)}{2}\\
   \includegraphics[scale=.15]{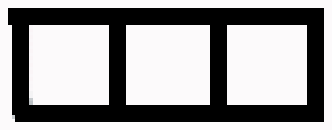}&\frac{N(N+1)(N+2)}{6}&\frac{(N+2)(N+3)}{2}&\frac{(N+3)(N+6)}{2}\\
  \includegraphics[scale=.15]{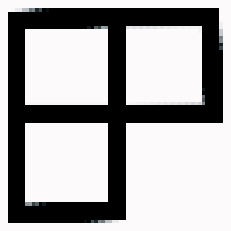}&\frac{N(N-1)(N+1)}{3}&N^2-3&N^{2}-9\\
   \includegraphics[scale=.15]{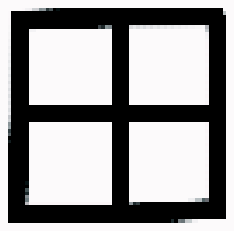}&\frac{N^{2}(N+1)(N-1)}{12}&\frac{N(N-2)(N+2)}{3}&\frac{N(N-4)(N+4)}{3}\\
   \includegraphics[scale=.15]{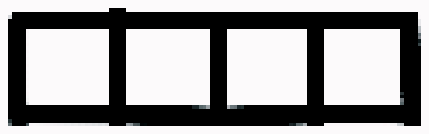}&\frac{N(N+1)(N+2)(N+3)}{24}&\frac{(N+2)(N+3)(N+4)}{6}&\frac{(N+3)(N+4)(N+8)}{6}\\
   \includegraphics[scale=.15]{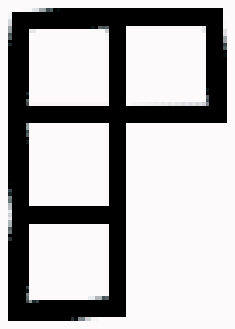}&\frac{N(N+1)(N-1)(N-2)}{8}&\frac{(N-2)(N^2-N-4)}{2}&\frac{(N-4)(N^{2}-N-8)}{2}\\
 \hline
\end{array}\]
\caption{Dimensions, Dynkin indexes, and anomaly coefficients for some
representations of SU(N).}
 \label{table:SUNreps1}
\end{table}

\begin{table}[ht]
\[\begin{array}{|c|c|c|c|}\hline
 Irrep & dim(r) & A(r)\\  \hline
    \includegraphics[scale=.15]{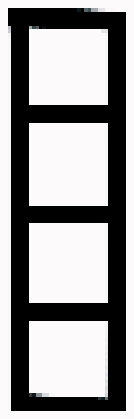}&\frac{N(N-1)(N-2)(N-3)}{24}&\frac{(N-4)(N-3)(N-8)}{6}\\
   \includegraphics[scale=.15]{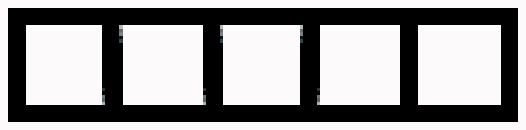}&\frac{N(N+1)(N+2)(N+3)(N+4)}{120}&\frac{(N+3)(N+4)(N+5)(N+10)}{24}\\
   \includegraphics[scale=.15]{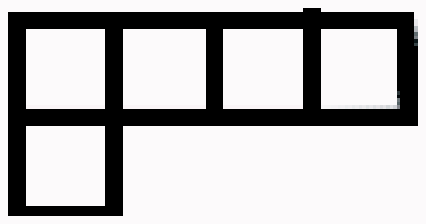}&\frac{N(N+1)(N+2)(N+3)(N-1)}{30}&\frac{(N-2)(N+3)(N+5)^{2}}{6}\\
   \includegraphics[scale=.15]{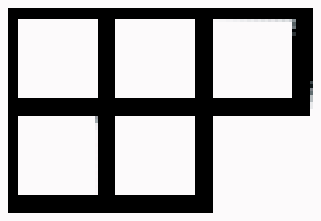}&\frac{N^{2}(N+1)(N+2)(N-1)}{24}&\frac{N(N+5)(5N^{2}-3N-50)}{24}\\
   \includegraphics[scale=.15]{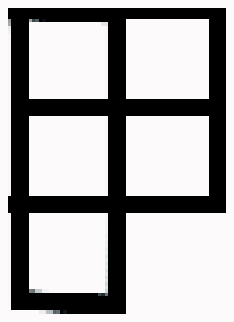}&\frac{N^{2}(N+1)(N-1)(N-2)}{24}&\frac{N(N-5)(5N^{2}+3N-50)}{24}\\
   \includegraphics[scale=.15]{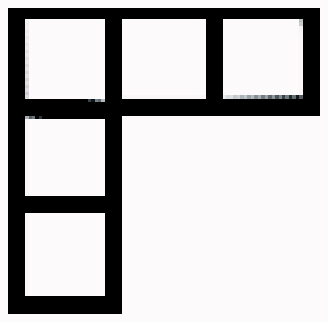}&\frac{N(N+1)(N+2)(N-1)(N-2)}{20}&\frac{(N^{4}-17N^{2}+100)}{4}\\
   \includegraphics[scale=.15]{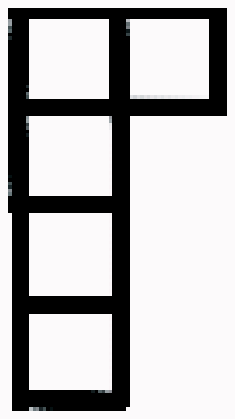}&\frac{(N-3)(N-2)(N-1)N(N+1)}{30}&\frac{(N-5)^{2}(N-3)(N+2)}{6}\\
   \includegraphics[scale=.15]{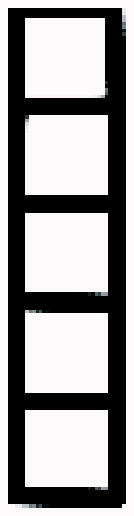}&\frac{N(N-1)(N-2)(N-3)(N-4)}{120}&\frac{(N-5)(N-4)(N-3)(N-10)}{24}\\
   \includegraphics[scale=.15]{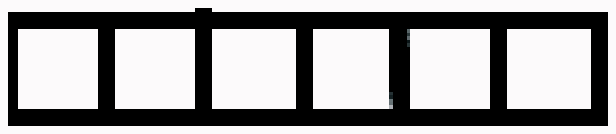}&\frac{N(N+1)(N+2)(N+3)(N+4)(N+5)}{720}&\frac{(N+3)(N+4)(N+5)(N+6)(N+12)}{120}\\
   \includegraphics[scale=.15]{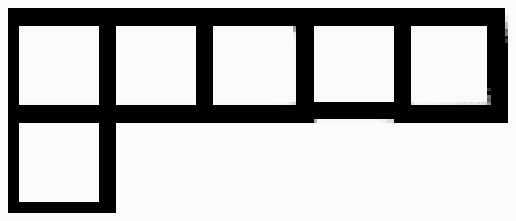}&\frac{N(N+1)(N+2)(N+3)(N+4)(N-1)}{144}&\frac{(N+3)(N+4)(N+6)(N^{2}+5N-12)}{24}\\
   \includegraphics[scale=.15]{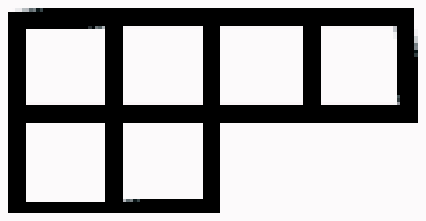}&\frac{(N-1)N^{2}(N+1)(N+2)(N+3)}{80}&\frac{3}{40}(N-3)N(N+3)(N+4)(N+6)\\
   \includegraphics[scale=.15]{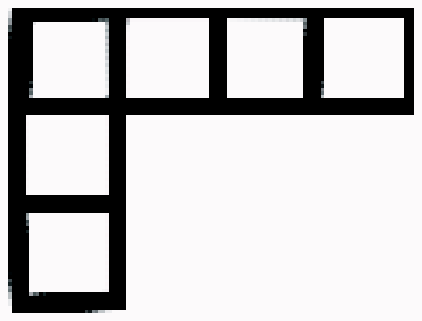}&\frac{(N-2)(N-1)N(N+1)(N+2)(N+3)}{72}&\frac{(N+3)(N^{4}+3N^{3}-16N^{2}-36N+144)}{12}\\
   \includegraphics[scale=.15]{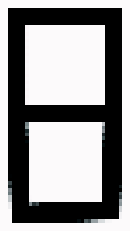}&\frac{(N-1)N^{2}(N+1)^{2}(N+2)}{144}&\frac{(N-4)N(N+1)(N+3)(N+6)}{24}\\
   \includegraphics[scale=.15]{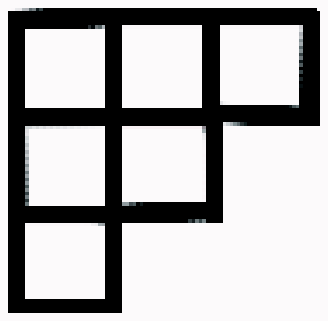}&\frac{(N-2)(N-1)N^{2}(N+1)(N+2)}{45}&\frac{2}{15}(N-4)(N-3)N(N+3)(N+4)\\
   \includegraphics[scale=.15]{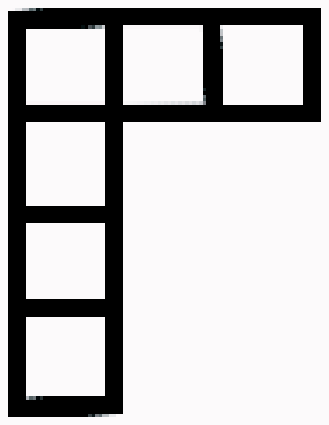}&\frac{(N-3)(N-2)(N-1)N(N+1)(N+2)}{72}&\frac{(N-3)(N^{4}-3N^{3}-16N^{2}+36N+144)}{12}\\
  \includegraphics[scale=.15]{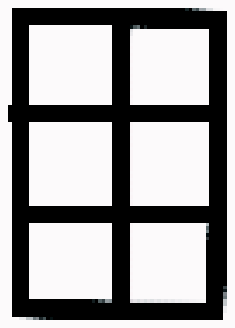} &\frac{(N-2)(N-1)^{2}N^{2}(N+1)}{144}&\frac{(N-6)(N-3)(N-1)N(N+4)}{24}\\
   \includegraphics[scale=.15]{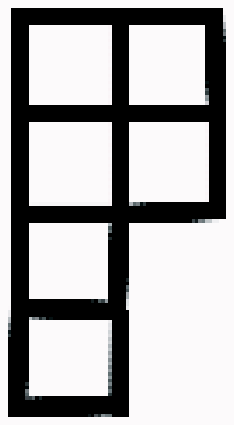}&\frac{(N-3)(N-2)(N-1)N^{2}(N+1)}{80}&\frac{3}{40}(N-6)(N-4)(N-3)N(N+3)\\
   \includegraphics[scale=.15]{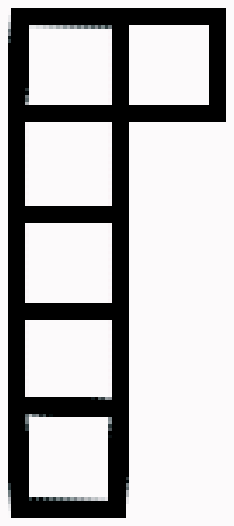}&\frac{(N-4)(N-3)(N-2)(N-1)N(N+1)}{144}&\frac{(N-6)(N-4)(N-3)(N^{2}-5N-12)}{24}\\
   \includegraphics[scale=.15]{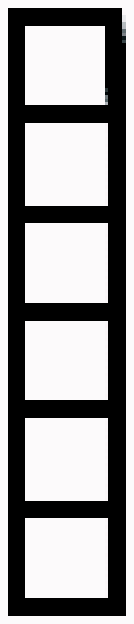}&\frac{(N-5)(N-4)(N-3)(N-2)(N-1)N}{720}&\frac{(N-6)(N-5)(N-4)(N-3)(N-12)}{120}\\
       \hline
\end{array}\]
\caption{Dimensions and anomaly coefficients for higher-dimensional
representations of SU(N).}
 \label{table:SUNreps2}
\end{table}

Then, given the previous results, one can try to identify
possible ways that will enable us  to construct  anomaly free
models. As it has been considered in the literature \cite{GeorgiGlashow72,BanksGeorgi},
there are several ways to obtain anomaly free theories, namely:

\begin{description}
\item[i)] The gauge group itself is safe, i.e. it is always free
of anomalies. This happens, for instance, for SO(10) but not for
SU(5).
\item[ii)] The gauge group is a subgroup of an anomaly free
  group, and the representations form a complete representation
  of the anomaly free group. For instance, this happens in the SU(5) case for the
  $\mathbf{5}+\mathbf{\overline{10}}$ representations, which together  are anomaly
  free, and this can be understood  because they belong to the $\mathbf{16}$ representation of
  SO(10), i.e. under SU(5) the \textbf{16} decomposes as: $\mathbf{16=5+\overline{10}+1}$.
\item[iii)] The fermionic representations appear in conjugate
  pairs, i.e., they are vector-like. This is the most common choice
  when the Higgs sector of SUSY GUT is extended \footnote{Although
    Higgs scalars do not contribute to the anomaly, in SUSY models
    they come with the Higgsinos, their fermionic partners, which can
    contribute to the anomaly.}.
  For instance, a
  $\mathbf{45}+\mathbf{\overline{45}}$ pair is considered to
  solve the problem associated with the wrong Yukawa unification for
  first and second families within SU(5) models.
\end{description}

Here, we  shall show that there are also other accidental
possibilities that result when several lower-dimensional Higgs
multiplets contribute  to the anomaly associated with a larger-dimensional 
Higgs representation. This will be illustrated with the
SU(5) case in the following section.

\section{Anomaly Cancellation in SUSY SU(5)}
Let us consider an SU(5) SUSY GUT model. There are three copies of
$\mathbf{\overline{5}}$ and $\textbf{10}$ representations to
accommodate the three families of quarks and leptons. Breaking of
the GUT group to the SM: $SU(5) \rightarrow SU(3)_C\times SU(2)_L
\times U(1)$, is achieved by including a (chiral) Higgs
supermultiplet in the adjoint representation (\textbf{24}).
Regarding anomalies, the $\mathbf{\overline{5}}$ and $\textbf{10}$
contributions cancel each other. This situation
corresponds to case ii in the previous section, that results
from the fact that the SU(5) gauge symmetry is a subgroup of  SO(10).
On the other hand, the \textbf{24} representation is itself
anomaly free. The minimal Higgs sector
needed to break the SM gauge group can be formed with a pair
of \textbf{5} and $\mathbf{\overline{5}}$ representations, which is
indeed vectorial and therefore anomaly free (this corresponds
to case iii discussed above).

Now, within this minimal model with a Higgs sector consisting of
\textbf{5}+$\mathbf{\overline{5}}$, one obtains the mass relations
$m_{d_i}=m_{e_i}$, which works well for the third family, 
but not for the second family,
while it may or may not work for the first family, depending on
whether or not one includes weak scale threshold
effects\cite{DMP}. One way to solve this problem is to add a
\textbf{45} representation, which couples to the d-type quarks
but not to the up-type, and one then obtains the Georgi-Jarlskog
factor~\cite{GeorgiJarlskog} needed for the correct mass
relations. Most models that obtain these relations with an
extended Higgs sector, include the conjugate representation
in order to cancel the anomalies, i.e.
$\mathbf{45}+\mathbf{\overline{45}}$ ~\cite{Pavel Fileviez}. 
This is however not the only possibility, and this is
one of the main results of our paper.

\begin{table}[ht]
  \[\begin{array}{|c|c|c|c|c|}\hline
  Irrep &Multiplet &dim(r) & A(r)&2T(r)\\
  \hline
  \left[5\right]&(0,0,0,0)&1&0&0\\
  \left[1\right]&(1,0,0,0)&5&1&1\\
  \left[2\right]&(0,1,0,0)&10&1&3\\
  \left[1,1\right]&(2,0,0,0)&15&9&7\\
  \left[4,1\right]&(1,0,0,1)&24&0&10\\
  \left[1,1,1\right]&(3,0,0,0)&35&44&28\\
  \left[2,1\right]&(1,1,0,0)&40&16&22\\
  \left[3,1\right]&(1,0,1,0)&45&6&24\\
  \left[2,2\right]&(0,2,0,0)&50&15&35\\
   \hline
  \end{array}\]
  \caption{Dimension, anomaly coefficients, and Dynkin indexes for different representations of SU(5).}
  \label{table:SU(5)}
\end{table}

The results for the anomaly coefficients for some representations
of SU(5) (and their conjugates) are shown
in table~\ref{table:SU(5)}; we can see that the \textbf{45}
anomaly coefficient is 6. Then taking into consideration that the
\textbf{5} and the \textbf{10} have the anomaly coefficient $A=1$, we can
write down the following anomaly-free combinations

\begin{eqnarray}
  A(\mathbf{45})+A(\mathbf{\overline{45}})=0& \ , \\
  A(\mathbf{45})+6A(\mathbf{\overline{5}})=0& \ , \\
  A(\mathbf{45})+6A(\mathbf{\overline{10}})=0& \ .
\end{eqnarray}
Alternatively we can write a general anomaly-free condition with these fields,
\begin{equation}
  A(\mathbf{45})+fA(\mathbf{\overline{5}})+f'A(\mathbf{\overline{10}})=0, \  \text{with  }f+f'=6 \ .
\end{equation}

One could also invoke a $\mathbf{\overline{15}}$ representation, which has
$A=-9$, through the following anomaly-free combination:
\begin{equation}\label{}
    A(\mathbf{45})+A(\mathbf{\overline{15}})+3A(\mathbf{5})=0 \ .
\end{equation}

These are non-equivalent models with different physical
consequences. This is explicitly shown in the next section
where we discuss the issue of gauge coupling unification.

\section{Gauge coupling unification and perturbative validity.}

The  $\beta$ functions for a general SUSY theory with
gauge group $G$ and matter fields appearing in chiral supermultiplets,
at the 1-loop level, are given by:
\begin{equation}\label{betafunctions}
\beta_1=\sum_R T_R - 3C_A,
\end{equation}
where $ T_R $ denotes the Dynkin index for the representation $R$,
and $C_A$ is the quadratic Casimir invariant for the adjoint
representation. For SU(N) type gauge groups $C_A=N$, while the
$T_R$ index for  most common SU(5) representations are also shown
in Table~\ref{table:SU(5)}.

The RGE's with 1-loop $\beta$ functions for the gauge couplings of
the MSSM are
%write 1-loop beta's and RGE

\begin{equation}\label{oneloopRGE}
    \frac{d\alpha_i}{dt}=\beta_i \alpha_i^2,
\end{equation}
where
\begin{equation}\label{oneloopBetas}
    \beta_i=\left(%
\begin{array}{c}
  33/5 \\
  1 \\
  -3 \\
\end{array}%
\right)+\beta^X
\end{equation}
and $t=(2\pi)^{-1}\ln M$, with $M =$ mass scale. The index $i=1,2,3$ refers to
the U(1), SU(2) and SU(3) gauge groups respectively. The term $ \beta^X=\sum_{\Phi}
T(\Phi)$ denotes the contributions of the additional representations 
beyond those included in the MSSM (the
sum is over all SU(5) additional multiplets $\Phi$). Assuming
$M_{SUSY}\approx M_{t}$, one obtains that the unified gauge coupling is
approximately $g(M_{GUT}) = 0.0416$, and unification
occurs at $M_{GUT}=2\times 10^{16}$~GeV~\cite{Hempfling}.

These simple 1-loop results can be improved by
using the 2-loop RGEs~\cite{Jones:1981we,Yamada:1993ga,Barger:1992ac}. 
In such case we solve numerically
the corresponding RGE and we find that at the GUT scale
$M_{GUT}=1.28\times 10^{16}$~GeV, the unified gauge coupling is
$g_5(M_{GUT}) = 0.040$, and $h^t(M_{GUT})=0.6572$.

Now we are interested in evaluating the effect of the different
representations in the running from $M_{GUT}$ up to the Planck
scale. Besides evaluating the effect of the different anomaly free
combinations, we are also interested in finding which
representations are perturbatively valid up to the Planck scale.
The unified gauge coupling obeys the 1-loop RGE
\begin{equation}
  \mu \frac{d \alpha_5^{-1}}{d\mu}=\frac{-\beta_1}{2\pi}=\frac{3-\beta^X}{2\pi} \ ,
\end{equation}
where $-\beta_1=\beta_{MIN}-\beta^X$, with $\beta_{MIN}=3 $
denoting the contribution to the SU(5) $\beta$~function
from the MSSM multiplets, including the one from the gauge sector.

The 1-loop $\beta$~functions for some interesting anomaly-free
combinations are found to be:
\begin{eqnarray}\label{betafuntions} \nonumber
    \beta^X(\mathbf{45}+\mathbf{\overline{45}} )& = & 24 \ , \\ \nonumber
    \beta^X(\mathbf{45}+ 6 (\mathbf{\overline{5}})) & = & 15  \ , \\  \nonumber
    \beta^X(\mathbf{45}+ 6 (\mathbf{\overline{10}})) & = & 21  \ , \\ \nonumber
    \beta^X(\mathbf{45}+ \mathbf{\overline{15}}+2 (\mathbf{10})+\mathbf{5}) & = & 19  \ , \\
    \beta^X(\mathbf{50}+\mathbf{\overline{40}}+\mathbf{5}) & = & 29 \ .
\end{eqnarray}

\begin{figure}[ht]
  \centering
  \includegraphics[width=14cm]{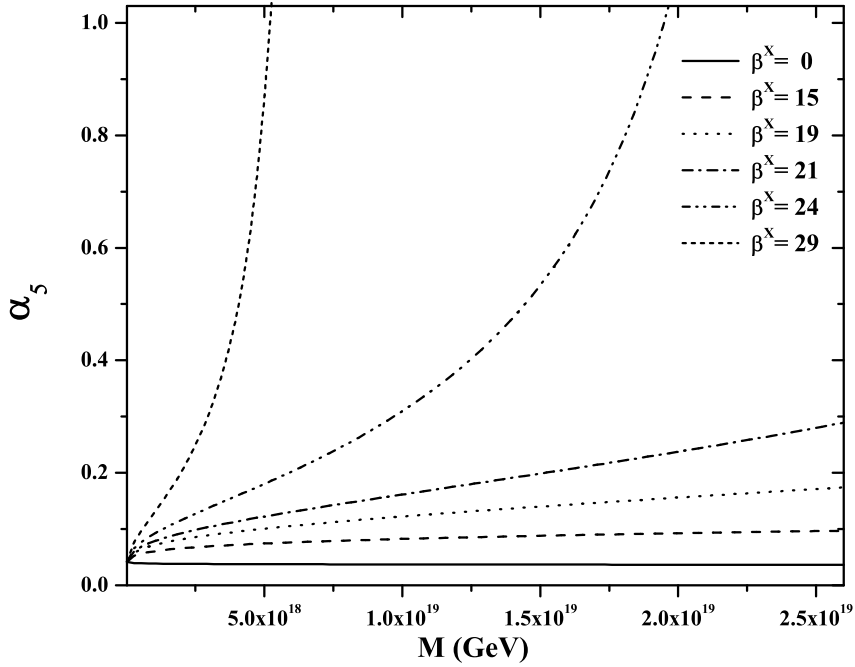}
  \caption{Evolution of the unified gauge coupling for the free
    anomaly combinations listed in the text. The evolution
    is shown all the way up to the Planck scale. }
  \label{fig:grunning1}
\end{figure}

As shown in Figure~\ref{fig:grunning1}, the
model with $\beta^X=29$ induces a running of the unified gauge  coupling
that blows at the scale $M=6.61\times 10^{18}$~GeV, while for
$\beta^X=24$ this happens at $M=2.63\times 10^{19}$~GeV. The models
with $\beta^X=15,19,21$ are found to evolve safely all the way up to the
Planck scale.

It is also interesting to consider the RGE effect associated with
the Yukawa coupling that involve the additional Higgs representations.
  In order to do this we shall consider the 2-loop $\beta$~functions
for the gauge coupling~\cite{Martin:1993zk}, but will keep only the 1-loop RGE for the new 
Yukawa couplings. Thus, we shall consider the following superpotential for the
SUSY SU(5) GUT  model: 

\begin{eqnarray}\label{} \nonumber
  W &=&  \frac{f}{3}Tr\Sigma^3+\frac{1}{2}f VTr\Sigma^2
  +\lambda\bar{H}_{\alpha}(\Sigma^{\alpha}_{\beta}+3V\delta^{\alpha}_{\beta})H^\beta \\
  &+&\frac{h^{ij}}{4} \varepsilon_{\alpha\beta\gamma\delta\epsilon}\psi^{\alpha\beta}_{i}\psi^{\gamma\delta}_{j}H^{\epsilon}
  +\sqrt{2}f^{ij}\psi^{\alpha\beta}_{i}\phi_{j\alpha}\bar{H}_{\beta} \ .
\end{eqnarray}

Note that this superpotential involves the Higgs representations
$\mathbf{5}$, $\bar{\mathbf{5}}$ y $\mathbf{24}$.

The 1-loop RGEs for the Yukawa parameters are given by~\cite{HMY93}:

\begin{eqnarray}\label{RGE SUSY SU(5)}
  \mu \frac{d\lambda}{d\mu} & = &
  \frac{1}{(4\pi)^2}\left(-\frac{98}{5}g_5^2+\frac{53}{10}\lambda^2
  +\frac{21}{40}f^2+3(h^t)^2\right)\lambda, \\
  \mu \frac{d f}{d\mu} & = &
  \frac{1}{(4\pi)^2}\left(-30g_5^2+\frac{3}{2}\lambda^2
  +\frac{63}{40}f^2\right)f, \\
  \mu \frac{d h^t}{d\mu} & = &
  \frac{1}{(4\pi)^2}\left(-\frac{96}{5}g_5^2+\frac{12}{5}\lambda^2
  +6(h^t)^2\right)h^t \ ,
\end{eqnarray}
while the 2-loop RGE for the unified gauge coupling is given by:
\begin{equation}\label{g5}
  \mu\frac{d
    g_5}{d\mu}=\frac{1}{(4\pi)^2}(-3g_5^{3})+\frac{1}{(4\pi)^4}\frac{794}{5}g_5^{5}-
                \frac{1}{(4\pi)^4}\left\{\frac{49}{5}\lambda^2
                 +\frac{21}{4}f^2+12(h^t)^2\right\}g^3 \ .
\end{equation}

\begin{figure}[ht]
  \centering
  \includegraphics[width=14cm]{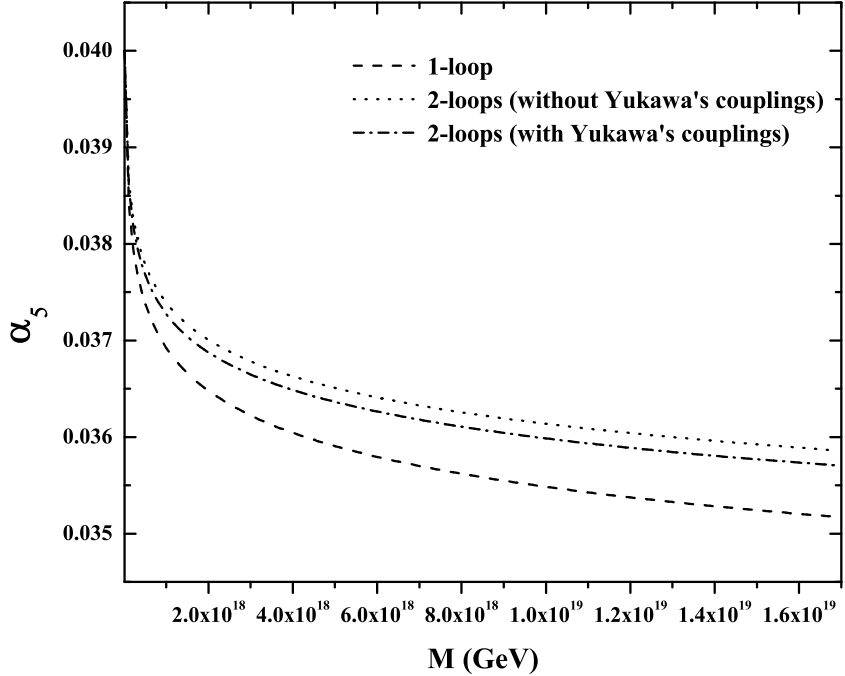}
  \caption{Evolution of the unified gauge coupling for three different cases:
    i) the 1-loop result, ii) the 2-loop result without including Yukawas, and iii)
    the 2-loop result including the (1-loop) running of the Yukawas.}
  \label{fig:grunning2}
\end{figure}

We use values of the coefficients $\lambda$, $h^t$ and $f$ that
are safe at the Planck scale, and look for their effects on the
unified gauge coupling. The resulting evolution is shown in one of the lines in
Figure~\ref{fig:grunning2}, where we show the 1-loop results, as well as the the 
2-loop results with and without the Yukawas 1-loop contributions. The parameters used in
the plots are $ M_{GUT}=1.28 \times 10^{16}$~GeV, $\alpha(M_{GUT})=0.040$,
$h^t(M_{GUT})=0.6572$, $\lambda(M_{GUT})=0.6024$, and $f(M_{GUT})=1.7210$. 
We notice that there are appreciable differences
for the evolution of the gauge coupling when going from the one to the 2-loop
cases, but this difference is reduced when one includes Yukawa couplings at
the 1-loop level.

\section{Conclusions}

We have studied the problem of anomalies in SUSY gauge theories
in order to search for alternatives to the usual vector-like
representations used in extended Higgs sectors. 
The known results have been extended to include higher-dimensional 
Higgs representations, which in turn have been applied to discuss
anomaly cancellation within the context of realistic GUT models of
SU(5) type. We have succeeded in
identifying ways to replace the $\mathbf{45}+\bar{\mathbf{45}}$
models within SU(5) SUSY GUTs. Then, we have studied the $\beta$~functions 
for all the alternatives, and we find that they are not
equivalent in terms of their values.
We have also considered the RGE effect associated with
the Yukawa coupling that involve the additional Higgs representations.
We found that there are appreciable differences
for the evolution of the gauge coupling when going from the 1 to the 2-loop
RGE, but this difference is reduced when one includes the 1-loop Yukawa couplings at
the 2-loop level.
These results have important implications for the perturbative
validity of the GUT models at scales higher than the unification
scale.

\begin{acknowledgments}
  This work was supported in part by CONACYT and SNI. A.A. acknowledges the
  Benem\'erita Universidad Aut\'onoma de Puebla for its warm hospitality while
  part of this work was being done.
\end{acknowledgments}

\end{document}